\documentclass[nofootinbib,prd,twocolumn,superscriptaddress]{revtex4}

\usepackage{empheq}
\usepackage{amsfonts}
\usepackage{soul}
\usepackage[normalem]{ulem}
\usepackage[english]{babel}
\usepackage[latin1]{inputenc}
\usepackage{lastpage} 
\usepackage[automark,komastyle]{scrpage2}
\usepackage{amssymb}
\usepackage{bm}
\usepackage{graphicx}
\usepackage[hang,normal]{subfigure}
\usepackage{enumerate}
\usepackage[usenames,dvipsnames]{color}
\usepackage[colorlinks=true,
		linkcolor=black,citecolor=black,
		urlcolor=black,breaklinks=true,bookmarksnumbered=true,
		hypertexnames=false,pdfpagemode=UseOutlines,
		plainpages=false,pdfpagelabels,bookmarks=true,
		linktocpage=false,pdfstartview=FitH]{hyperref}
\hypersetup{pdfauthor={Christine Gruber},pdftitle={},
		pdfsubject={},pdfcreator={pdfLaTeX with hyperref (\today)}}

\newcommand{\figref}[1]{Fig.~\ref{#1}}

\newcommand{\eqeqref}[1]{Eq.~\eqref{#1}}
\newcommand{\eqsref}[1]{Eqs.~\eqref{#1}}
\newcommand{\secref}[1]{Section~\ref{#1}}

\def\be{\begin{equation}}
\def\ee{\end{equation}}
\def\bea{\begin{eqnarray}}
\def\eea{\end{eqnarray}}

\allowdisplaybreaks

\begin{document}
\title{An effective model for inflation from geometrothermodynamics: a detailed analysis of thermodynamics and cosmological 
perturbations}
\date{\today}

\author{Christine Gruber}
\email{christine.gruber@correo.nucleares.unam.mx}
\affiliation{Instituto de Ciencias Nucleares, Universidad Nacional Aut\'onoma de M\'exico,\\ AP 70543, M\'exico, 
DF 04510, Mexico.}

\author{Hernando Quevedo}
\email{quevedo@nucleares.unam.mx}
\affiliation{Instituto de Ciencias Nucleares, Universidad Nacional Aut\'onoma de M\'exico,\\ AP 70543, Ciudad de M\'exico 04510, 
Mexico;\\
Dipartimento di Fisica and ICRANet, \\
Universit\`a di Roma ``La Sapienza", \\ I-00185 Roma, Italy; \\
	Department of Theoretical and Nuclear Physics,  \\ Kazakh National University, \\Almaty 050040, Kazakhstan}

\begin{abstract}
Inflationary models usually assume the existence of scalar fields or other exotic gravitational sources. 
We investigate an alternative approach in which the entire Universe is considered as a thermodynamic system
described by geometrothermodynamics. A particular van der Waals like fundamental equation is used to construct an 
effective inflationary model which is consistent with the main physical requirements,  such as the number of e-foldings 
and the times for the onset and end of inflation, predicting in this way a volume of approximately $10^{-84}$ m$^3$ at the onset of
inflation. The phase transition structure and thermodynamic behavior of the system are consistent with the expected properties of
an inflationary scenario. Cosmological perturbations of the model are shown to be in agreement with the corresponding primordial 
power spectrum providing the seeds for the creation of large-scale structure.
\end{abstract}

\maketitle

\section{Introduction}

Currently the most accepted general framework describing the evolution of the Universe is the $\Lambda$CDM 
model. Based on general relativity, it postulates the existence of an inflationary era in the early development 
of the Universe, which can be well approximated by a cosmological constant dominating over all other components 
of the Universe at that time, such as baryonic matter, dark matter or radiation \cite{2002Riot}. These different 
constituents, usually modeled by fluids, are each described by their own equations of state, describing their 
expansion behavior. The way to fix the 
equation of state is usually by choice, but it can also be obtained from more profound thermodynamic principles. 
As already once put to use in \cite{abcq12}, and in a previous work by the authors \cite{2017Grub}, there exists a method to derive  equations of state 
from a fundamental thermodynamic equation, within a formalism called geometrothermodynamics (GTD) \cite{quev07}. 

GTD, providing a differential geometric description of thermodynamics, states that thermodynamic systems of any kind 
can be described by a corresponding geometrical space of equilibrium states, denoted by a manifold ${\cal E}$. The 
geometrical properties of this manifold encode the thermodynamic behavior of the system in equilibrium. For instance, the 
geometrical curvature of the equilibrium manifold can be understood as an indication and a measure of 
the thermodynamic interactions of the system, and possible curvature singularities correspond to critical points 
in thermodynamics such as phase transitions. Many confirmations of these principles have been found in various 
thermodynamic systems (see \cite{qqs16b} and references therein). 

For systems with $n$ thermodynamic degrees of freedom, i.e., $n$ independent thermodynamic variables describing 
its behavior, the corresponding equilibrium manifold ${\cal E}$ is $n-$dimensional as well. Its curvature and 
other geometric properties are obtained from an $n-$dimensional metric $g$ on ${\cal E}$, which is determined 
by the thermodynamic properties of the system. Usually, the metric $g$ and its components are assumed to be the 
second derivatives of a thermodynamic potential describing the system. Depending on the approach, this potential 
can be the internal energy (Weinhold metric), the entropy (Ruppeiner metric), or can be calculated from a partition 
function (Fisher-Rao metric). All of these metrics have been used in various contexts and in the description of 
many thermodynamic systems and their statistical theories \cite{weibook,rup14,amari85}. GTD however does not use 
these Hessian metrics, but instead defines $g$ by requiring it to be Legendre invariant. Legendre invariance, i.e., 
the independence of the thermodynamic behavior of a system on the choice of thermodynamic potential, is a central 
property of thermodynamics, and should be reflected in the elements of its description. The incorporation of 
Legendre transformations as coordinate transformations in GTD requires the introduction of a higher-dimensional 
structure to embed the equilibrium manifold, i.e., a  $(2n+1)-$dimensional phase space ${\cal T}$ equipped with 
a contact 1-form $\Theta$ and a  $(2n+1)-$dimensional metric $G$. 
The coordinates on this manifold are $Z^A=\{\Phi, E^a, I^a\}$,
with $a=1,2, ..., n$,  
which are interpreted as the thermodynamic variables describing 
a system, and the 1-form and the metric can be expressed in 
terms of these coordinates. On the equilibrium manifold ${\cal E}$,  
$\Phi$ is interpreted 
as a thermodynamic potential, the $E^a$ as the extensive variables 
and the $I_a$ as the intensive variables dual to the $E^a$. 
The phase space ${\cal T}$ and the equilibrium space ${\cal E}$ 
are connected by an embedding map 
$\varphi: {\cal E} \rightarrow {\cal T}$, its pullback 
satisfying the condition $\varphi^*(\Theta) = 0$. When fixing 
the coordinates on ${\cal E}$ as the $E^a$, the pullback 
condition becomes the first law of thermodynamics. The metric 
on ${\cal E}$ is given as a function of the $E^a$, and is 
determined by fixing a dependence $\Phi (E^a)$, i.e., 
by choosing how the thermodynamic potential depends on the 
extensive variables. 

Usually in classical thermodynamics, empirical 
methods based on laboratory experiments are used to 
find the fundamental equation for each system \cite{callen}. 
GTD provides an alternative way based on properties of the 
geometrical structures describing the system \cite{vqs10}. 
If we suppose  that the equilibrium manifold ${\cal E}$ is an 
extremal subspace of ${\cal T}$, with its action 
$I=\int \sqrt{|{\rm det}( g) |}\, d^n E$ 
satisfying the variational principle $\delta I =0$, then we 
obtain the Nambu-Goto differential equations 
\be
  \frac{1}{\sqrt{|{\rm det}( g) |}}\left(\sqrt{|{\rm det}( g) |}\,\, g^{ab}Z^A_{,a}\right)_{,b} + 
  \Gamma^A_{\ BC} Z^B_{,b}Z^C_{,c} g^{bc} =0 \,,
  \label{meqg}
\ee
with $\Gamma^A_{\ BC}$ being the Christoffel symbols associated 
with the metric $G$. The system of differential equations 
\eqref{meqg} determines possible solutions for $Z^A(E^a)$, and 
thus for the fundamental equation $\Phi(E^a)$. Assuming a 
particular choice $\Phi = S$ and $E^a=(U,V)$, with $S$ as the 
entropy, $U$ the internal energy and $V$ the volume of the 
system, \eqsref{meqg} can be used to determine different fundamental potentials. 
A particularly simple solution that extremizes the Nambu-Goto action can be written as
\be
  S = c_1 \ln \left( U + \frac{\alpha}{V} \right) + c_2 \ln \left( V - \beta  \right) \,,
  \label{gvdw}
\ee
where $c_1$, $c_2$, $\alpha$ and $\beta $ are real constants. The thermodynamic variables are the internal energy $U$, 
the volume $V$, the temperature $T$ and the pressure $P$ of the fluid. Choosing $c_1 = 2/3$, $c_2=1$, $\alpha>0$ and $\beta >0$ 
leads to the fundamental equation of the van der Waals gas. However, the solution \eqref{gvdw} admits more general forms 
of the fundamental equation, with different values for the constants, which are not fixed by the Nambu-Goto equations 
\eqref{meqg}. 

Applying the analysis of GTD to this case, it turns out that in general the equilibrium space corresponding to \eqref{gvdw} is curved, 
which signifies that the system contains thermodynamic interactions. In the limit of $\alpha=\beta=0$, $c_1=2/3$ and $c_2=1$, the ideal gas is 
recovered, which features no thermodynamic interactions, and thus the curvature of its equilibrium space metric $g$ 
vanishes. In general, different equations of state of ideal fluids can be obtained with different choices of $c_1$ and $c_2$, which 
have been successfully applied in cosmology. In the present work,
we expand upon the work of a previous article \cite{2017Grub}, 
where we investigate the behavior and physical consequences 
of this generalized interacting van der Waals type fundamental 
equation in the context of cosmology. In \cite{2017Grub}, we assume 
the presence of such a fluid in the early stages of the Universe, 
and found it capable of describing an inflationary period at the 
beginning of the Universe's evolution. In this article, we 
present details on the thermodynamic properties of the inflationary fluid, 
and investigate the cosmological perturbations generated by it, 
in order to evaluate its viability as an effective inflationary 
model. 

This paper is structured as follows. 
In \secref{sec:yestd}, we review the findings of \cite{2017Grub} 
and present the solution for the fluid density generating 
inflation, briefly discussing its properties and parameters. 
The thermodynamic properties of the fluid are thoroughly 
investigated in \secref{sec:thbehav}, including an analysis 
of the evolution of the thermodynamic variables, the response 
functions and possible phase transitions and critical points. 
In \secref{sec:pert}, we comment on the primordial perturbations  
generated from the inflationary fluid, which provide the seeds of 
large-scale structure formation in the subsequent eras of cosmological 
evolution. In \secref{sec:sum}, we discuss our results and mention some tasks for 
further investigation.  

Throughout the paper we use geometric units with 
$G=c=\hbar=k_{_B}=1$.

\section{Cosmological behavior and properties of the inflationary fluid}
\label{sec:yestd} 

In this section, we review the findings of \cite{2017Grub}, 
where the fundamental equation \eqref{gvdw} of a fluid with interactions has been found and applied in the context of 
the early Universe. \eqeqref{gvdw} was used to derive the 
equations of state of the fluid, 
\begin{equation}
  \frac{1}{T} = \frac{\partial S}{\partial U} \quad \mathrm{~and~} \quad \frac{P}{T} = \frac{\partial S}{\partial V} \,,
\end{equation}
resulting in 
\begin{equation} \label{eq:temp}
  T = \frac{U}{c_1} + \frac{\alpha}{c_1 V} \,
\end{equation}
and 
\begin{equation} \label{eq:pressureorig}
  P = \frac{c_2 U V^2 + \alpha \left[ \beta c_1 + \left( c_2 - c_1 \right) V \right] }{c_1 V^2 \left( V - \beta \right)} \,.
\end{equation}

The energy density was defined via the ratio of the internal 
energy and the volume as 
\begin{equation} \label{eq:density}
  \rho = \frac{U}{V} \,.
\end{equation}
Note that this definition of energy density is fundamentally different from other works in the literature in this direction 
\cite{2016Jant,2017Vard}, where a van der Waals gas has been parameterized in terms of a density 
$n = N/V$. In those approaches, the equation of state is $P=P(V,T)$, i.e., the pressure as a function of the temperature 
and the volume, using \eqref{eq:temp} to substitute $U$ in \eqref{eq:pressureorig}. The temperature then enters as a 
free parameter into these models, and is usually fixed as an arbitrary constant throughout inflation. In our case, we 
use \eqref{eq:pressureorig}, the pressure as a function of 
the internal energy and the volume, and keep both variables free. We thus do not have to specify a temperature 
during inflation, and obtain a model with much larger generality, in which the temperature is obtained as an evolving 
variable from the dynamics of the corresponding thermodynamic system. The results thus differ significantly from previously obtained solutions. 

Besides introducing the density \eqref{eq:density}, we parametrize the volume as a function of the scale factor $a(t)$ as 
\begin{equation} \label{eq:Vofa}
  V = V_0 a^3 \,.
\end{equation}
Here, we are using the conventional choice of $a(t_0) = 1$, where $t_0$ is the current time, implying that $V_0$ 
is the volume of the Universe at current time. At earlier times, the scale factor is thus always smaller than unity. 
With this, the pressure as a function of the density and the scale factor is obtained as 
\begin{equation} \label{eq:Pofa}
  P = \frac{a^9 c_2 \rho V_0^3+\alpha \left[a^3 V_0 (c_2-c_1)+\beta c_1\right]}{a^6 c_1 V_0^2 \left(a^3 V_0-\beta\right)} \,.
\end{equation}
Inserting this expression into the continuity equation for a cosmic fluid 
evolving in an FLRW-background 
leads to a differential equation for the density $\rho$ as a function of the scale factor $a$. It is solved to give 
\begin{equation} \label{eq:rho}
  \rho = \frac{K \left(a^3 V_0 - \beta\right)^{-\frac{c_2}{c_1}} }{a^3}-\frac{\alpha}{a^6 V_0^2} \,,
\end{equation}
with $K$ being an integration constant. This is the most 
general dynamical behavior of the energy density of the 
fluid obtained for the fundamental equation \eqref{gvdw}. 
As shown in \cite{2017Grub}, if we fix $c_2/c_1 = -8/9$, the density 
in \eqref{eq:rho} is the exact expression for the density producing inflation. Using approximations we can get a qualitative picture 
of how inflation is achieved. Indeed, since $V=V_0a^3$ represents the volume of the Universe during inflation, the constant $\beta$ 
in \eqref{gvdw} can be interpreted as the volume at the beginning 
of inflation, which is expected to become negligible compared 
to the term $V_0a^3$ very fast. We can therefore expand
the density for small $\beta$, resulting in 
\begin{equation} \label{eq:rhofirstapprox}
  \rho \simeq \frac{K V_0^{8/9}}{a^{1/3}} - \frac{8 \beta K}{9 V_0^{1/9} a^{10/3}} - \frac{\alpha}{a^6 V_0^2} \,.
\end{equation}
The first term proportional to $a^{-1/3}$ has been shown to produce the required 
amount of expansion for an inflationary period, provided
the constants $\alpha$ and $\beta$ are sufficiently small. 
As demonstrated in \cite{2017Grub}, the optimal way to fix them 
is by identifying 
\begin{equation} \label{eq:condInfStrong}
 \left| \alpha \right| = \frac{8}{9} \beta K V_0^{17/9} a_i^{8/3} \,,
\end{equation}
with $\alpha < 0$, where $t_i$ is the time of the onset of 
inflation, and $a_i = a(t_i)$ the corresponding scale factor. 
Further, we introduce a smallness parameter, 
\begin{equation} \label{eq:epsilon}
  \epsilon = \frac{\beta}{\beta_c} = \frac{|\alpha|}{\alpha_c} \,, 
\end{equation}
with 
\begin{equation}
	\beta_c = \frac{9}{8}\, V_0 \,a_i^3 \,,
\  
 \left| \alpha_c \right| = \frac{8}{9} \beta_c K V_0^{17/9} a_i^{8/3} \,.
\end{equation}
Assuming such conditions on $\alpha$ and $\beta$, we can calculate the amount of expansion during inflation generated 
by the inflationary density 
\begin{equation} \label{eq:rhoinf}
  \rho_{\mathrm{inf}}\equiv \frac{K V_0^{8/9}}{a^{1/3}} \,.
\end{equation}
It turns out that for an inflationary period lasting from 
an initial time $t_i = 10^{-36}$ until a final time $t_f = 10^{-32}$ \cite{2002Riot}, the inflationary density can 
generate about $N \simeq 55$ e-foldings of expansion, an 
appropriate amount according to standard inflationary 
theories \cite{2002Riot,2003Line}. Furthermore, considering standard assumptions on the history 
of the Universe and its expansion \cite{2017Grub,2003Line},  
we obtain 
\be
  \beta_c = \frac{9}{8} V_0 a_i^3 \simeq 10^{-84} \, \mathrm{m^3} \,,
\ee
i.e., a small number, about the volume of the Universe at the 
beginning of inflation, but larger than the Planck volume. 
Similarly, estimating the bound on $\alpha$ requires an estimate 
on $K$, which we fix using the condition that inflation begins roughly at the GUT energy scale. $K$ is thus found to be 
\begin{equation}
   K = \frac{\rho_{\mathrm{inf}} a_i^{1/3}}{V_0^{8/9}} \simeq 2\cdot 10^{2} \, \mathrm{J m^{-17/3}} \,.
\end{equation}
With this, the critical value of $\alpha$ then results to be 
\begin{equation}
  \alpha_c \simeq 10^{-78} \, \mathrm{J m^3} \,.
\end{equation}

Having specified the inflationary density and all its 
parameters and properties from a dynamical point of view, 
we can now proceed to investigate its thermodynamic 
properties and behavior.

\section{Analysis of thermodynamic behavior}
\label{sec:thbehav}

In this section, we will discuss the evolution of temperature, 
pressure and response functions of the inflationary fluid, 
followed by an analysis of its phase diagram, calculating 
its phase transition structure and critical point.

\subsection{Temperature}

The temperature is given by \eqref{eq:temp}, and can 
be rewritten using \eqref{eq:Vofa}, \eqref{eq:Pofa} and 
\eqref{eq:rho} as 
\begin{equation} \label{eq:Texact}
  T = \frac{K}{c_1} V_0 \left( a^3 V_0 - \beta \right)^{-\frac{c_2}{c_1}} \,.
\end{equation}
Using the inflationary value $c_2/c_1 = -8/9$, and expanding the expression for small 
$\beta$, the temperature becomes 
\begin{equation} 
  T \simeq  \frac{K}{c_1}  a^{8/3} \left( V_0^{17/9} - \frac{8 \beta  V_0^{8/9}}{9} \frac{1}{a^3} \right) \,.
\end{equation}
With \eqref{eq:epsilon} for $\epsilon$, and introducing the 
dimensionless variable 
\begin{equation} \label{eq:x}
  a(t) = a_i x \,,
\end{equation}
the temperature can be represented  as 
\begin{equation} \label{eq:Tinf}
   T \simeq  \frac{K}{c_1} V_0^{17/9} a_i^{8/3} x^{8/3} \left( 1 - \epsilon x^{-3} \right) \,.
\end{equation}
We assume without loss of generality $c_1>0$. 
The temperature is positive if the expression in the 
bracket is positive, i.e. as long as $\epsilon$ 
is small at the onset of inflation. Thus, a small $\epsilon$ 
not only produces an inflationary period, but also guarantees 
that the temperature is a positive quantity. For $x<1$, the 
temperature can become negative, but in the regime before 
the onset of inflation the equation of state is assumed not 
to be valid. 

\begin{figure}[h]
    \centering
    \includegraphics[width=.49\textwidth]	
    {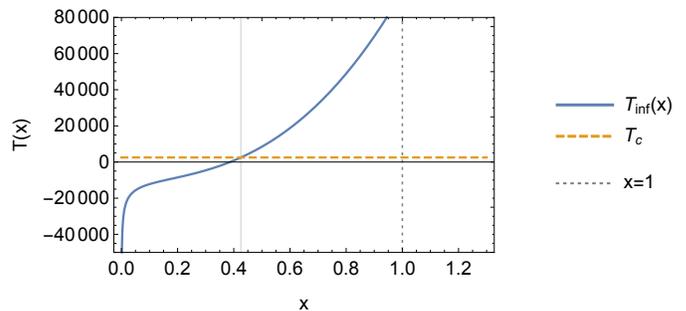} 
    \caption{The figure shows the evolution of the temperature 
    (solid blue) throughout the inflationary period, valid 
    starting from $x=1$. We see that around $x_{c,T}\simeq 
    0.425$ the critical temperature (dashed orange) is reached, 
    and that slightly before that, the temperature becomes 
    negative, and thus unphysical.}
    \label{fig:T}
\end{figure}

These qualitative features can be seen in \figref{fig:T}, 
where we have plotted the temperature, as given in 
\eqref{eq:Texact}, as a function of $x$, introducing the 
smallness parameter $\epsilon$ in place of $\alpha$ and 
$\beta$. We see that in the inflationary period, starting from 
$x=1$, the temperature is positive and increasing. It reaches 
the critical temperature at $x_{c,T}\simeq 0.425$, and 
is negative before $x \simeq 0.402$. This  unphysical region 
lies already outside of the supposed regime of validity of 
the inflationary equation of state. For this plot we fix the 
smallness parameter as $\epsilon = 0.065$ because, as we will 
see in \secref{sec:pert}, this value guarantees an appropriate 
power spectrum of cosmological perturbations. We thus use this 
value for $\epsilon$ in all further expressions with numerical 
values and curves.

\subsection{Pressure}

The pressure in terms of the scale factor is given, using \eqref{eq:Pofa} and \eqref{eq:rho} with 
$c_2/c_1 = -8/9$, as 
\begin{equation} \label{eq:Pressofa}
  P = \frac{9 a^3 \alpha V_0 + 8 a^6 K V_0^3 \left(a^3 V_0-\beta\right)^{8/9} - 9 \alpha \beta}
	{9 a^6 \beta V_0^2-9 a^9 V_0^3} \,.
\end{equation}
The expansion for small $\beta$ leads to 
\begin{equation} 
  P \simeq -\frac{\alpha}{a^6 V_0^2} - \frac{8 K}{9} V_0^{8/9} a^{-1/3} 
      - \frac{8 \beta K}{81} V_0^{-1/9} a^{-10/3} \,.
\end{equation}
Again substituting the scale factor by $x = a / a_i$, and 
introducing  $\epsilon$, the pressure can be written as 
\begin{equation} \label{eq:Pinfofx}
  P_{inf}(x) = \frac{\rho_{\mathrm{inf}}(a_i)}{x^{1/3}} \left[ - \frac{8}{9} - \frac{\epsilon}{9} x^{-3} 
    + \epsilon x^{-17/3} \right] \,.
\end{equation}
With the growth of $x$ during inflation, the factor $x^{-1/3}$ decreases, and the pressure thus evolves from its 
initial value 
\begin{equation} 
  P(x_i) = -\frac{8}{9} \rho_{\mathrm{inf}} (a_i) \left( 1 - \epsilon \right) \,
\end{equation}
to less negative values. In the limit of small $\epsilon$, the pressure is negative and nearly that of a cosmological 
constant, with a barotropic factor $\omega_i = - 8 /9$. 

\begin{figure}[h]
    \centering
    \includegraphics[width=.49\textwidth]	
    {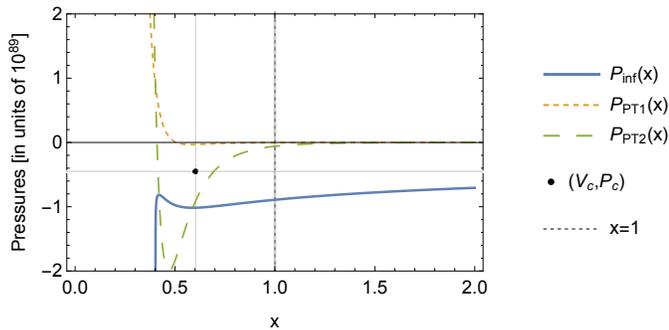} 
    \caption{The phase diagram shows the pressure as a 
    function of $x$ (or the volume, respectively). The 
    inflationary dynamics are depicted in the  solid 
    curve, while the short-dashed  and long-dashed 
     curves denote the two kinds of phase transitions 
    in the system. We see that the inflationary evolution 
    crosses the second phase transition (long-dashed curve) 
    twice, which means there are two phase transitions before 
    the onset of inflation. The black dot denotes the critical 
    values of volume and pressure, and the dotted line 
    the onset of inflation at $x=1$.}
    \label{fig:PhaseDiag}
\end{figure}

We show the evolution of the exact inflationary 
pressure, as given by \eqeqref{eq:Pressofa}, in the phase 
diagram  shown in \figref{fig:PhaseDiag},  where we use the variable $x$ and the smallness parameter $\epsilon$ instead of the 
scale factor and the constants $\alpha$ and $\beta$, respectively.
The pressure as a function of $x$ is equally a 
time-evolution plot, since $x$ grows during inflation 
starting at $x=1$, as well as the phase diagram, 
since the volume grows monotonously with the variable $x$. 
Indeed, the typical van der Waals like curve is recovered, 
in this case inverted to negative pressures, due to the 
negativity of the parameters $\alpha$ and $c_2$. The 
behavior shown in the phase diagram is in accordance to 
the approximate expression \eqref{eq:Pinfofx}, which has 
been derived to qualitatively understand the evolution of 
the pressure during inflation and its asymptotic behavior. 

In contrast to a usual van der Waals phase diagram, here we 
do not have different curves for different temperatures. The 
timelike evolution of the temperature during the inflationary 
period is fixed by the dynamics of the fluid within a 
FLRW-universe. There is thus only one fixed curve for the 
evolution of the pressure, shown in \figref{fig:PhaseDiag}. 
Again, we have chosen $\epsilon = 0.065$ to plot the curve, 
in accordance with the findings of \secref{sec:pert}.

\subsection{Response functions}

The response functions of the system are the speed of sound $c_S$, the heat capacities at constant volume $c_V$ and constant pressure 
$c_P$, the isothermal compressibility $\kappa_T$, the isentropic compressibility $\kappa_S$ and the thermal expansion $\alpha_V$. 

The adiabatic speed of sound is defined as 
\begin{equation}
  c_S^2 = \frac{\partial P}{\partial \rho} = \frac{\partial P}{\partial a} \left( \frac{\partial \rho}{\partial a} \right)^{-1} \,,
\end{equation}
and can be calculated from $P(a)$ and $\rho(a)$, resulting in 
a very long and cumbersome expression. We will recast it later 
in terms of $x$ and $\epsilon$, in order to qualitatively assess 
its properties. 

The heat capacity at constant volume is given by 
\begin{equation}
  c_V = \left( \frac{\partial U}{\partial T} \right)_V = c_1 \,.
\end{equation}
In order to have a positive temperature, we have assumed that $c_1>0$, and thus also $c_V>0$. 
That means that $c_2$ is negative since $c_2 / c_1 = -8/9$ holds. A positive heat 
capacity at constant volume is thus quite reasonable, 
just as in a conventional van der Waals fluid. 

The isothermal compressibility is defined as 
\begin{equation}
  \kappa_T = -\frac{1}{V} \left( \frac{\partial V}{\partial P} \right)_T \,,
\end{equation}
which can be calculated as 
\begin{equation} \label{eq:compT}
  \kappa_T = \frac{V^2 (V-\beta)}{2 \alpha \beta-\alpha V+P V^3} \,.
\end{equation}
The isentropic compressibility is 
\begin{equation}
  \kappa_S = -\frac{1}{V} \left( \frac{\partial V}{\partial P} \right)_S \,,
\end{equation}
which can be calculated, substituting the temperature as a function of the entropy for the derivation, as 
\begin{equation} \label{eq:compS}
 \kappa_S = \frac{c_1 V^2 (V-\beta)}{\alpha \left[ 2 \beta c_1+V (c_2-c_1) \right] +P V^3 (c_2+c_1)} \,.
\end{equation}
~\\
The thermal expansion coefficient is defined as 
\begin{equation} \label{eq:alphaV}
  \alpha_V = \frac{1}{V} \left( \frac{\partial V}{\partial T} \right)_P = \kappa_T \left( \frac{\partial P}{\partial T} \right)_V \,,
\end{equation}
leading to 
\begin{equation}
  \alpha_V = \frac{c_2 V^2}{2 \alpha \beta-\alpha V+P V^3} \,.
\end{equation}
~\\
The heat capacity at constant pressure is defined as 
\begin{equation} \label{eq:cP}
  c_P = \left( \frac{\partial H}{\partial T} \right)_V = c_V + V T \frac{\alpha_V^2}{\kappa_T} \,,
\end{equation}
and can be calculated from the other response functions as 
\begin{equation} \label{eq:cPcalc}
  c_P = c_1 + \frac{c_2 V \left(\alpha+P V^2\right)}{2 \alpha \beta-\alpha V+P V^3} \,.
\end{equation}

In our model 
$\alpha$ and $c_2$ are negative, and thus the pressure 
results to be completely inverted to negative pressures. 
This implies that in the isothermal compressibility 
\eqref{eq:compT} the denominator changes sign, and the 
compressibility becomes negative, as opposed to the usual 
van der Waals fluid with  positive $\alpha$ and $c_2$. 
Also in the isentropic compressibility, the factors of 
$\alpha$ and pressure $P$ in the denominator lead to a 
sign change. Moreover, negative values of $c_2$ lead to 
changes in the divergences of the quantity, due to the 
factors of $(c_2 \pm c_1)$ in the denominator. The 
compressibilities for normal fluids are positive, i.e. the volume decreases with increasing pressure, and the higher 
the pressure, the harder it gets to compress the fluid. 
Here, the effect is reversed - which means that with 
increasing pressure, the volume increases as well, something 
that intuitively corresponds well to an inflationary 
behavior and a fluid with negative pressure. 

In the thermal expansion \eqref{eq:alphaV}, both numerator 
and denominator change sign, and thus the quantity remains 
of the same sign as for a van der Waals fluid. Usually the 
thermal expansion is positive, i.e. materials expand with 
increasing temperature, if the pressure is held constant. 
The temperature of the expanding universe is thus increasing 
along with the expansion, driven by the inflationary pressure. 
This indicates that there is no need for a reheating period in 
this model, since the thermodynamic properties of the fluid 
do not lead to a cooling of the universe during the inflationary 
expansion, as in conventional models. 

Also the heat capacity at constant pressure could possibly 
change its behavior, depending on the value of $c_2$. 
Rewriting expression \eqref{eq:cPcalc} as 
\begin{equation} 
  \frac{c_P}{c_1} - 1 = \frac{c_2 V \left(\alpha+P V^2\right)}{2 \alpha \beta-\alpha V+P V^3} \,,
\end{equation}
we see that the fraction on the right-hand side remains the 
same, due to sign changes in numerator as well as denominator, 
and that its sign is determined by the ratio $c_2/c_1$. 
Depending on the magnitude of $c_2/c_1$ and the evolution 
of the fraction, the heat capacity at constant pressure can 
become negative. 

In order to get a more quantitative grip on these analysis, 
also the response functions can be converted from 
functions  in terms of $P$ and $V$ into functions of 
$x$ and in terms of $\epsilon$, as 
\begin{widetext}
\begin{eqnarray} 
 && c_S^2(x) \simeq - \left[ \frac{4}{18} x^{17/3} - \epsilon \right] \left[ \frac{1}{18} x^{17/3} + \epsilon \right]^{-1} 
    - \frac{40}{9^2} \frac{\epsilon}{x^3} \left[ \frac{1}{16} x^{17/3} - \epsilon \right] \left[ \frac{1}{18} x^{17/3} + \epsilon \right]^{-2} 
		\,, \label{eq:cs2xe} \\[8pt]
 && \kappa_T(x) \simeq \frac{1}{\rho_{\mathrm{inf}}(a_i)} \left[ \frac{4}{9} x^{17/3} - \epsilon \right]^{-1} \left[ -\frac{x^6}{2} + 
		\frac{5}{18} \frac{\epsilon x^{26/3}}{\left( \frac{4}{9} x^{17/3} - \epsilon \right)} \right] \,, \\[8pt]
 && \kappa_S(x) \simeq \frac{1}{\rho_{\mathrm{inf}}(a_i)} \left[ \frac{4}{81} x^{17/3} - \epsilon \right]^{-1} \left[ -\frac{x^6}{2} + 
		\frac{5}{162} \frac{\epsilon x^{26/3}}{\left( \frac{4}{81} x^{17/3} - \epsilon \right)} \right] \,, \\[8pt]
 && \alpha_V(x) \simeq \frac{32 c_1 \beta_c}{81 \alpha_c} \left[ \frac{4}{9} x^{26/3} - \epsilon x^3 - \frac{\epsilon}{18} x^{17/3} 
		- \frac{9}{8} \epsilon^2 \right] \left[ \frac{4}{9} x^{17/3} - \epsilon \right]^{-2} \,, \\[8pt]
 && \frac{c_P(x)}{c_1} \simeq \left[ \frac{4}{9} x^{17/3} - \epsilon \right]^{-1} \left[ \frac{4}{81} x^{17/3} - \epsilon + 
    \frac{40}{81} \epsilon^2 x^{-1/3} \left[ \frac{4}{9} x^{17/3} - \epsilon \right]^{-1} \right] \,. 
\end{eqnarray}
\end{widetext}
The coefficient factor in the compressibilities can be expressed, using the inflationary density evaluated at the onset of inflation, as 
\begin{equation} 
  \rho _{\mathrm{inf}}(x_i) = \frac{K V_0^{8/9}}{a_i^{1/3}} =: \rho_{GUT} \,.
\end{equation}
In the limit of small $\epsilon$, the response functions behave like 
\begin{eqnarray}
  c_S^2 &\sim& -4 < 0 \,, \\[8pt]
  \kappa_T &\sim& -\frac{9}{8} \frac{x^{1/3}}{\rho_{GUT}} < 0 \,, \\[8pt]
  \kappa_S &\sim& -\frac{81}{8} \frac{x^{1/3}}{\rho_{GUT}} < 0 \,, \\[8pt]
  \alpha_V &\sim& \frac{72 c_1 \beta_c}{\alpha_c} x^{-8/3} > 0\,, \\[8pt]
  \frac{c_P}{c_1} &\sim& \frac{1}{9} >0 \,. 
\end{eqnarray}
Here we see that the speed of sound is imaginary and from \eqref{eq:cs2xe} we deduce that 
for growing $x$ it remains negative. 
This is unphysical, but not uncommon for inflationary 
models. The above-mentioned behavior of the compressibilities 
as well as the thermal expansion is confirmed here, the compressibilities becoming negative and the thermal expansion remaining positive. 
The heat capacity turns out to be positive as well, as for 
conventional fluids, and remains so during the course of 
inflation with growing $x$.

\subsection{Phase transitions, critical point and phase diagram} 

According to GTD \cite{quev07,qqs16b}, phase transitions of 
a thermodynamic system described by the fundamental equation (\ref{gvdw})  are given by the solutions of the 
equation  
\begin{equation} \label{eq:PTs}
  P V^3 - \alpha V + 2 \alpha \beta = 0 \,.
\end{equation}
In order to depict these solutions in a phase diagram, we 
express the pressure as a function of the volume along the 
phase transition rearranging \eqeqref{eq:PTs} as 
\begin{equation} \label{eq:PT1}
  P_{PT1} = \frac{\alpha V - 2 \alpha \beta}{V^3} \,.
\end{equation}
Using the parametrization of the volume in terms of the 
variable $x$ as given in \eqeqref{eq:Vofa}, we can obtain the pressure at 
phase transition in dependence of $x$, again for the chosen 
value of $\epsilon = 0.065$. This yields the first phase 
transition line in \figref{fig:PhaseDiag}, corresponding to the  
short-dashed curve. 

Moreover, we have to check the response functions for points 
in which they are not well-behaved, for instance,  when their denominators 
show zeros. Most of the response functions, namely the heat 
capacities, the compressibility at constant temperature and 
the thermal expansion, have the previously mentioned equation 
\eqref{eq:PTs} as their denominator, i.e. they misbehave 
exactly at the points already specified by \eqref{eq:PTs}. 
Only the compressibility at constant entropy differs slightly, 
with its denominator reading 
\begin{equation} \label{eq:PTkappa}
  P V^3 - 17 \alpha V + 18 \alpha \beta = 0 \,.
\end{equation}
Repeating the same operations as above, this equation can be transformed to yield a second phase transition at 
\begin{equation} \label{eq:PT2}
  P_{PT2} = \frac{17\alpha V - 18 \alpha \beta}{V^3} \,.
\end{equation}
This phase transition is depicted as the long-dashed 
line in the phase diagram \figref{fig:PhaseDiag}, as a 
function of $x$ for $\epsilon = 0.065$. We see that the 
inflationary pressure stays clear of the phase transition 
pressure $P_{PT1}$, but crosses the line of the phase 
transition pressure $P_{PT2}$ twice, implying two phase 
transitions indicated by a divergent isentropic 
compressibility near the beginning of inflation, at 
$x_{1} \simeq 0.42$ and $x_2 \simeq 0.589$, the second 
phase transition being very close to the value of the critical volume, 
as we will see in the following. 

In order to calculate the critical point, we analyze the pressure 
\begin{equation}
  P = \frac{c_2 T}{V-\beta} - \frac{\alpha}{V^2} \,,
\end{equation}
obtained from combining \eqsref{eq:temp} and \eqref{eq:pressureorig}. Reformulating this relation into a 
polynomial of the volume, we end up with 
\begin{equation}
  V^3 + V^2 \left( \frac{-\beta P - c_2 T}{P} \right) + \frac{\alpha V}{P} - \frac{\alpha\beta}{P} = 0 \,.
\end{equation}
Identifying this expression with the generic polynomial 
\begin{equation}
  (V- V_c)^3 = V^3 +-3 V_c V^2 + 3 V_c^2 V - V_c^3 = 0 \,,
\end{equation}
the critical point $(V_c,P_c,T_c)$ is calculated as 
\begin{equation} \label{eq:CritPoint}
  V_c = 3 \beta \,, \quad P_c = \frac{\alpha}{27 \beta^2} \,, \quad T_c = \frac{8\alpha}{27 \beta c_2} \,.
\end{equation}
We now want to know where the critical point is situated in 
the phase diagram, and whether it is reached during the course 
of inflation. 

The volume in its evolution during inflation can be parameterized 
in terms of $x$ as 
\begin{equation}
  V(x) = V_0 a(t_i)^3 x^3 = \frac{8}{9} \beta_c x^3 \,.
\end{equation}
By equating the volume and the critical volume, we can calculate the critical value, i.e., 
\begin{equation} \label{eq:VequalVc}
	V(x_{c,V}) = \frac{8}{9} \beta_c x_{c,V}^3 = 3 \beta = V_c \,.
\end{equation}
Resolving this equation, fixing $\epsilon=0.065$, leads to 
\begin{equation} \label{eq:xcforV}
	x_{c,V} = \frac{3}{2} \epsilon^{1/3} = 0.603 \,,
\end{equation}
i.e., the critical volume is reached slightly before the onset 
of inflation. 

The critical pressure can be expressed in terms of $\epsilon$ as well, leading to 
\begin{equation} \label{eq:Pc}
  P_c = - \frac{1}{\epsilon} \frac{2^6}{3^7} \rho_{\mathrm{inf}} (a_i) \,.
\end{equation}
Using $\epsilon=0.065$, the critical pressure can be 
calculated, which permits us to mark the  point $(V_c,P_c)$ 
in the phase diagram in \figref{fig:PhaseDiag} as a black dot. 
We see thus that the universe does not cross over 
the critical point in this inflationary model, evolving along 
$P_{inf} (x)$. 

Ultimately, we have to analyze the critical temperature, given 
by the third expression in \eqeqref{eq:CritPoint}. It can be 
expressed in terms of the parameters of the inflationary 
model as 
\begin{equation} \label{eq:Tc}
  T_c = \frac{1}{3} \frac{K}{c_1} a_i^{8/3} V_0^{17/9} \,,
\end{equation}
where the sign of $\alpha$ has been canceled by a factor of $c_2/c_1$ occurring in the expression. Comparing the critical 
temperature with the exact inflationary temperature from 
\eqeqref{eq:Texact}, expressed in terms of the variable $x$, 
we can calculate when the critical temperature will be reached by equating $T(x_{c,T}) = T_c$. The result is, assuming 
$\epsilon = 0.065$, 
\begin{equation} \label{eq:xcforT}
  x_{c,T} = 0.425 \,.
\end{equation}
The critical temperature is thus reached before inflation as 
well, and the inflationary dynamics, valid from $x=1$, do 
not cross it.

\section{Cosmological Perturbations}
\label{sec:pert}

In this section, we investigate the cosmological perturbations generated by the thermodynamic system described by the
fundamental equation (\ref{gvdw}), and compare them to the results expected from an inflationary model. 
In analogy to \cite{abcq12} and \cite{1995MaBe}, 
we use the perturbed Einstein equations in the fluid limit, in the longitudinal (or conformal Newtonian) gauge, 
and consider perturbations to the energy density and pressure of the fluid. For the moment, we do not assume 
anything about the origin of these perturbations - they should be determined by microscopic processes following 
the statistical properties of the system, and thus require a thorough statistical analysis of the thermodynamic system 
presented here. In the present work, we are mainly interested in the cosmological consequences of the model. 

It is easy to show that in the thermodynamic system \eqref{gvdw} 
shear and anisotropic stresses are absent. 
Hence the corresponding energy-momentum tensor can be represented 
as that of a perfect fluid, 
\begin{equation}
	T^{\mu}_{~\nu} = P \delta^{\mu}_{~\nu} + \left( \rho + P \right) u^{\mu} u_{\nu} \,,
\end{equation}
where $u^{\mu}$ is the four-velocity of the fluid. Perturbations are introduced in this energy-momentum tensor 
by perturbing the density as $\rho = \bar{\rho} + \delta \rho$ and the pressure as $P = \bar{P} + \delta P$, 
where the barred quantities refer to the average values of the quantities. In contrast to \cite{1995MaBe}, we 
will not consider a peculiar fluid velocity or anisotropic shear perturbations. The perturbed energy momentum 
tensor thus reads 
\begin{equation}
	T^{0}_{~0} = - \left(\bar{\rho} + \delta \rho\right) \,, T^{0}_{~i} = 0 \,, T^{i}_{~j} = \left(\bar{P} + 
		\delta P \right) \delta^{i}_{j} \,.
\end{equation}

From the conservation of the perturbed energy-momentum tensor $T^{\mu \nu}_{~~;\mu} =0$, 
considering an equation of state $P=\omega \rho$ and a FLRW background
with scalar perturbations in the conformal Newtonian gauge 
\cite{1995MaBe} and in conformal time $\eta$, 
\begin{equation}
	ds^2 = a^2(\eta) \left[ -\left( 1 + 2 \Phi \right) d\eta^2 + \left( 1 - 2 \Phi \right) dx^i dx_i\right] \,,
\end{equation}
we obtain the equation determining the density contrast $\delta = \delta\rho/\bar{\rho}$ as 
\begin{equation} \label{eq:deltaprelim}
  \delta' = 3\left( 1+\omega \right) \Phi' - 3 \mathcal H \left( c_S^2 - \omega \right) \delta \,, 
\end{equation}
where prime denotes derivative wrt. conformal time $\eta$,  $c_S^2$ is the adiabatic speed of sound, and $\Phi$ is 
the metric perturbation in the conformal Newtonian gauge, i.e., the Newtonian gravitational potential determined 
as \cite{2002Riot,2016Krem}
\begin{equation}
  \Phi = 1 - \frac{\mathcal H}{a^2} \int a^2 d\eta \,. 
\end{equation}

To determine the density contrast $\delta$, we use $a(t)$ as 
in \cite{2017Grub} and convert it to conformal time to obtain 
$\mathcal H (\eta)$ and $a(\eta)$, and can thus calculate 
\begin{eqnarray}
  && \omega_i = \frac{P}{\rho_{inf}} = - \frac{8}{9} + \frac{8\epsilon}{9} -\frac{5 \epsilon^2}{72} \,, \label{eq:omega} \\
  && c_S^2 = - \frac{\left( \frac{4}{18} -\epsilon \right)}{\left( \frac{1}{18} +\epsilon \right)} - 
    \frac{40}{9^2} \frac{\left( \frac{1}{16} - \epsilon \right) \epsilon}{\left( \frac{1}{18} + \epsilon \right)^2} 
    + \mathcal O (\epsilon^2)\,, \label{eq:cs2} \\
  && \mathcal H = -\frac{6}{5\eta} \,,\\
  && \Phi (\eta) = 1 - \frac{6}{7} = \frac{1}{7} \,.
\end{eqnarray}
Note that the Newtonian potential is a constant. This does not 
mean that the primordial power spectrum is flat, but just that 
the gravitational potential introduced as a perturbation parameter 
in the metric coefficients is a constant throughout inflation. 
The equation for the classical density contrast  \eqref{eq:deltaprelim}, formulated in terms of the scale factor 
$a$ as variable instead of the conformal time, thus reads 
\begin{equation} \label{eq:deltaa}
  \frac{\partial \delta}{\partial a} = - \frac{3 \left( c_S^2 - \omega \right)}{a} \delta(a) \,.
\end{equation}
In terms of the dimensionless variable $x = a/a_i$ this equation 
becomes 
\begin{equation} 
  \frac{\partial \delta}{\partial x} = - \frac{3 \left( c_S^2 - \omega \right)}{x} \delta(x) \,.
\end{equation}
The general solution to this equation is 
\begin{equation}
  \delta(x) = \delta_0 x^{-3 \left( c_S^2 - \omega \right)} \,.
\end{equation}
In order to investigate the scale dependence of this function, we convert the dependence on $x$ into an inverse scale, or wavenumber, 
defined via $  k = \frac{2\pi}{x}$, resulting in 
\begin{equation} \label{eq:deltak}
  \delta(k) = \delta_{0} \, (2\pi)^{-3 \left( c_S^2 - \omega \right)} \, k^{3 \left( c_S^2 - \omega \right)}\,. 
\end{equation}
To compare the prediction for the inflationary spectrum and its tilt to what is expected from observations, 
we rewrite $\delta(k)$ in the form \cite{2002Riot}
\begin{equation}
	\mathcal{P}_{\mathcal{R}} (k) = \mathcal{A}^2_{\mathcal{R}} \left( \frac{k}{aH} \right)^{n-1} \,,
\end{equation}
where $n$ is the spectral index of the inflationary perturbation spectrum, $a$ the scale factor and $H$ 
the Hubble parameter. Casting \eqref{eq:deltak} in a similar form, we can thus identify 
\begin{eqnarray}
	\mathcal{A}^2_{\mathcal{R}} &=& \delta_{0} \, \left(\frac{a H}{2\pi}\right)^{3 \left( c_S^2 - \omega \right)} \,, \\
	n-1 &=& 3 \left( c_S^2 - \omega \right) \,,
\end{eqnarray}
The amplitude factor can be chosen to its desired value by fixing $\delta_{0}$. The qualitative aspect of the spectrum, 
i.e., the spectral tilt $n$ can be determined by using the dependence of the equation of state parameter \eqref{eq:omega} 
and the speed of sound \eqref{eq:cs2} on the small interaction 
parameter $\epsilon$. Defining 
\begin{equation}
	n (\epsilon) = 3 \left[ c_S^2 (\epsilon) - 
		\omega(\epsilon) \right] -1 \,,
\end{equation}
and solving $n (\epsilon) = 0.96$ as commonly expected 
\cite{2003Line}, we obtain 
\begin{equation}
\epsilon = 0.065\,,
\end{equation}
which is the value used in previous sections for investigating the thermodynamic properties of our inflationary model. 

Thus, it is possible to obtain a nearly scale-invariant spectrum 
for inflation by adjusting $\epsilon$ accordingly. We have shown 
that our model is viable for the description of inflation including 
the correct prediction of the inflationary power spectrum.

\section{Discussion \& Conclusions}
\label{sec:sum}

This work presents a phenomenological model for the period of inflation originating from the GTD formalism, a purely 
thermodynamic approach providing us with a thermodynamic fundamental equation, which is used in the context of 
relativistic cosmology. The fundamental equation used is the entropy of a thermodynamic system, parameterized in terms of its internal 
energy and its volume. It 
contains the typical van der Waals interaction parameters $\alpha$ and $\beta$, and two further constants, $c_1$ and 
$c_2$, generalizing the van der Waals case. 

Using this fundamental equation in the context of cosmology, it is a powerful model that can be applied to describe 
many epochs in the evolution of the Universe, depending on the choice of the involved parameters. For the limit of 
zero thermodynamic interactions, i.e., $\alpha=\beta=0$, the fluid successfully reproduces the behavior of the 
cosmological eras of radiation dominance, matter dominance and accelerated late-time expansion, by an appropriate 
choice of the constants $c_1$ and $c_2$. The interacting case has been the main focus of this article, and has been shown 
\cite{2017Grub} to provide a realistic model for the period of inflation in the early Universe. With the choice $c_2/c_1=-8/9$, 
it is possible to obtain predictions that are consistent with the main requirements for inflation, such as an amount 
of expansion of about $55$ e-foldings, and the times of onset and end of inflation. To obtain these features, 
limitations have to be placed on the magnitude of the interaction parameters $\alpha$ and $\beta$. The interaction 
strength $\alpha$ turns out to be negative, i.e., it corresponds to a repulsive interaction between the system's
constituents, and is bound by the value $\alpha_c = 10^{-78} \,\mathrm{J m^3}$. The parameter $\beta$ is 
found to be related to the size of the Universe at the onset of inflation, and is limited by $\beta_c = 10^{-84}
\, \mathrm{m^3}$, a value  which can be interpreted as a prediction of our model.  
The remaining parameters $V_0$ and $K$ have been fixed to physically reasonable values expected 
from inflationary models. Under these assumptions, the interacting thermodynamic system described by the fundamental equation
(\ref{gvdw}) has been shown to correctly mimic an 
inflationary period. 

By relating the two van der Waals parameters $\alpha$ and $\beta$ we have reduced the number of free parameters of 
the model further, and obtained the dynamical evolution of the fluid in terms of only one parameter, $\epsilon$, 
which is supposed to be small. The value of $\epsilon$ can be constrained by the consequences for cosmological 
perturbations predicted by the model. Even though there is a small freedom in the choice of the equation of state 
parameter, the resulting number of e-foldings is quite sensitive with respect to variations of its value, and so 
we consider the possible parameter space as well constrained. The model is thus unique, and not a part of a family 
of models with similar predictions, and therefore can be tested against observational evidence. 

Besides its cosmological and dynamical behavior, the thermodynamic properties of the model have been investigated. 
After calculating the temperature and the pressure, the response functions such as speed of sound, heat capacities, 
compressibilities and thermal expansion are obtained. The system shows conventional behavior of 
the heat capacities and thermal expansion, whereas its speed of sound is imaginary -- not uncommon for inflationary 
fluids -- and its compressibilities intuitively correspond to a fluid with repulsive interactions. We find that the critical 
point is not reached in the range of validity of the model. The analysis of the phase transitions 
however leads to interesting results, providing motivation for the inclusion of interaction in the system. It turns 
out that two phase transitions happen very close to the onset of inflation, interpreted as triggering the start of 
inflation. 
As an aside, we would like to comment that the introduction of interactions to the phases of radiation, matter and dark 
energy dominance leads to a similar occurrence of phase transitions at the beginning of these phases as in the case of 
inflation. These phase transitions could thus provide the general rationale for the change of eras in the Universe's 
history. 

Since the predictions of this model are unique, it is compelling to test its predictions against observations. In 
the case of an inflationary model, this can be done by calculating the form of the primordial perturbations the model 
generates. Our analyses demonstrate that the power spectrum of the primordial perturbations can be tuned to be nearly 
scale-invariant as expected by observations, by an appropriate choice of the smallness parameter $\epsilon$, thereby fixing the 
remaining semi-free parameter of the model. 

We have thus shown the success of an interacting van der Waals type fluid as the inflationary agent, without the 
necessity of postulating scalar fields or exotic mechanisms, solely relying on thermodynamic principles. As a next 
step, we would like to investigate the microscopic dynamics involved, utilizing analogies to the well-known van der 
Waals case, and delve deeper into the microscopic origins of the interactions and properties of the fluid, with the 
hope of improving the analysis of perturbations from a thermodynamic point of view.

\section*{Acknowledgements}
CG was supported by an UNAM postdoctoral fellowship program. This work has been supported by the
UNAM-DGAPA-PAPIIT, Grant No. IN111617.


\end{document}